\documentclass[preprint,prb,aps,showkeys,amsmath,amssymb]{revtex4-1}
\usepackage{epsfig}
\usepackage{amsmath}

\usepackage{color}

\begin{document}
\title{ Water: one molecule, two surfaces, one mistake   }
\author{  Carlos Vega }
\affiliation{Departamento de Qu\'{\i}mica F\'{\i}sica,
Facultad de Ciencias Qu\'{\i}micas, Universidad Complutense de Madrid,
28040 Madrid, Spain}
\date{\today}

C.Vega,  Mol.Phys. , {\bf 113}, 1145, (2015) \\

\begin{abstract}

 In order to rigorously evaluate the energy and dipole moment of a certain configuration of molecules
 one needs  to solve the Schr\"odinger equation. Repeating this for many different configurations
 allows one to determine the potential energy surface (PES) and the dipole moment surface (DMS).
 Since the early days of computer simulation it has been implicitly accepted that for empirical potentials the
 charges used to fit the PES should also be used to describe the DMS. This is a mistake.
 Partial charges are not observable magnitudes. They  should be regarded as adjustable
 fitting parameters. Optimal values used to describe the PES are not necessarily the best to describe the DMS.
 One could use two fits: one for the PES, and another for the DMS.
 This is a  common practice  in the quantum chemistry community,  but
 not used so often by the community performing computer simulations. 
 This idea affects all types of modelling of water (with the
 exception of ab-initio calculations) from coarse grained to non-polarizable and polarizable models.
 We anticipate that an area that will benefit dramatically from having both, a good PES and a good DMS, is
 the modelling of water in the presence of electric fields.

\end{abstract}

\maketitle


\section{Introduction}

Water is a simple molecule: just two hydrogens and one oxygen. Still it has a fascinating behavior related
 to the possibility of forming tetrahedral hydrogen bonded network 
 structures.\cite{bookStruPropWater,ball01,speedy76,poole92,water_structure_review,MP_2008_106_2053,review_structure_water_2014} The hydrogen bond, a  
 directional and rather strong intermolecular interaction (when compared to van der Waals  forces), is responsible 
 for  the special properties of water. Moreover the hydrogen 
 atoms are light  so  nuclear quantum effects are important. 
  Understanding the properties of water from a molecular point of view is certainly important.
 Computer simulations can be useful for that purpose, and they started with the seminal 
 papers of Barker and Watts\cite{barker69} in 1969 and of Rahman and Stillinger\cite{rahman71} in 1971. 
 Since the seminal paper of Bernal and Fowler\cite{bernal33,finney_revisa_bernal} 
 water is often described by a  Lennard-Jones (LJ) center  and several charges.
 The model of Bernal and Fowler was modified by Jorgensen et al.\cite{jorgensen83} to obtain 
 the popular TIP4P model. 
 Abascal and Vega have shown that the parameters of the TIP4P can be modified to yield a new model,
 TIP4P/2005.\cite{abascal05b} TIP4P/2005 is 
 a rigid non-polarizable model and one may wonder how far can one go in the description of water
 with such a simple model. Recently we have calculated a number of properties for this model and 
 compared them to experimental results.\cite{vega09} The comparison has been extended to other popular rigid
 non-polarizable water models such as SPC/E\cite{berendsen87}, TIP3P\cite{jorgensen83} and TIP5P\cite{mahoney00}. These are also rigid non-polarizable 
 models and they differ from TIP4P/2005 in the way the partial charges have been arranged.\cite{vega09,vega_perspective} 
 The comparison revealed some useful information. Not all water models are equally successful 
 in describing the experimental properties. From the considered models, TIP4P/2005 provided  
 the best results. However, since the model is rigid and non-polarizable it  can 
 not describe all the experimental properties of water. Thus, our feeling is that
 TIP4P/2005 represents the limit of the description of water that can be achieved by using rigid
 non-polarizable models. It is a decent model but to 
 go beyond that, new physical features (and not simply new parameters sets) must be incorporated. 
 
   We found a property with a somewhat surprising behavior: the
 dielectric constant. We found two puzzles when considering the dielectric constant of water. Firstly,
 certain models were able to describe the dielectric constant of water at room temperature and pressure.
 This is the case of TIP3P\cite{jorgensen83} and  TIP5P\cite{mahoney00}. However for some other models the dielectric constant was low when
 compared to experiment increasing in the order TIP4P, TIP4P/2005, SPC \cite{SPC} and  SPC/E\cite{berendsen87}. Secondly
 Rick and co-workers\cite{rick03,rick05,JCP_132_054509,rick11_cojounudo_ctes_dielectricas_hielos}, Lindberg and Wang\cite{dielectric_constant_ice} and 
 ourselves\cite{macdowell10,aragones11a,aragones11b} computed the dielectric constant of ice Ih. The surprising result was  
 that for this phase all these water models predicted a dielectric constant value lower than the experimental one, sometimes 
 by a factor of two. 
 The first reaction to explain these results is to assign the discrepancy to the approximate description 
 of the intermolecular potential. This is reasonable but still
 this hypothesis  should explain why all models fail in describing the dielectric constant of ice Ih.   
  Experimentally,  the dielectric constant of ice 
 Ih and water at the melting point are quite similar,  that of the solid phase being slightly higher.\cite{petrenko99} The importance of this
 experimental finding may not have been fully appreciated, and may be quite relevant, since it may 
 affect the way we approach the modeling of water. 
  At the melting point the tetrahedral order of liquid water 
 is quite high (that would explain the maximum in density) and it is even difficult to find order parameters (required
 for nucleation studies\cite{ARPC_2004_55_333,quigley08} ) 
 that distinguish between liquid-like  and a solid-like arrangements around a central 
molecule.\cite{trout03,conde08,geiger_dellago,jacs2013} 
 If  ice Ih and water are quite similar at the melting point, it is difficult to explain why all models 
 fail in describing the dielectric constant of ice Ih. 
  We found that for TIP4P models the dielectric constant of ice Ih was similar to that
 of liquid water (and this is in agreement with experiment) whereas the predicted value was too low for both phases  
 when compared to experiments.\cite{aragones11a} At this point we proposed in 2011 an explanation as to why TIP4P/2005 was unable to reproduce the dielectric 
 constant of water related to the failure of the model to describe the "real" water dipole moment 
 in condensed phases. Not surprisingly the title of our 2011 paper was " The dielectric constant of water and ices  a lesson about water interactions".\cite{aragones11a}
  This is probably true
 but in this paper we will present some evidence illustrating that maybe we did not obtain the ultimate 
 consequences of the "lesson". 
   
\section{ ABCD in the  modelling of water. }

 Let us consider a system with N molecules of water. Since each water molecule has three atoms, we need 
to define the position of the 3N atoms of the system, ${\bf R^{3N}}$.  $E^{0} ( {\bf R^{3N}}  )$ (which defines
the potential energy surface, PES) is the energy
of the system. We shall define the intermolecular potential energy U as:

\begin{equation}
\label{energia_potencial}
  U  ( {\bf R^{3N}}  )      = E^{0} ( {\bf R^{3N}}  ) -  N E^{0}_{H_{2}O}.
\end{equation}

where we have taken as zero of energies the energy of a system of N isolated
water molecules ($N E^{0}_{H_{2}O}$). The superscript zero indicates that there is no electric field present.
 
 It is useful for pedagogical  reasons to classify the different approaches in the modelling of water into four groups (or teams) 
which we will label as A, B, C and D. 
 They differ in the way U is obtained. 
 In Table \ref{equipos} the main four treatments in the modelling of water are presented.

 If you solve the Schr\"odinger equation to obtain $E^{0}$, then your treatment is of type A or B. 
In group A the motion of the nuclei is also treated from a quantum perspective. In group B one uses 
classical statistical Mechanics to describe the motion of the nuclei on the PES ( i.e 
the nuclei are regarded as classical objects). Approach B is often denoted as Car-Parrinello 
simulation\cite{PRL_1985_55_002471} and approach A as "full quantum". 
\textcolor{black}{ Within classical statistical mechanics the positions of the nuclei are governed by:}

\begin{equation}
-\nabla_{{\bf R_i} }(E^{0}( {\bf R^{3N} } )) =m_i\dfrac{ d^2 {\bf R_{i} } }{dt^2}
\end{equation}

\begin{equation}
 p( {\bf R^{3N} } ) \propto e^{-\beta E^{0}({\bf R^{3N}} )} 
\end{equation}

where the first expression (Newton's law) is to be used in Molecular Dynamics simulations and the second one in Monte Carlo (MC)  simulations,
being $ p( {\bf R^{3N} } )$ the probability of having a certain configuration of the nuclei. 
The approach A is described in Ref.\cite{full_quantum} and some examples for water 
within  the approach B can be found in Ref.\cite{JCP_2005_123_014501,artacho_water}. Notice that in approaches A and B, the energy
is obtained "on the fly"  for each configuration either by solving the Schr\"odinger equation or by performing density functional theory (DFT) calculations. 

\begin{table}
\caption{ Different approaches in the modeling of water }
\label{equipos}
\begin{tabular}{ccccc}
\hline \hline
  &   A    &    B    &   C    &    D    \\
\hline \hline
  &Electronic             &      Electronic        &     \textcolor{black}{Analytical}          &  \textcolor{black}{Analytical}               \\
Electrons  &structure              &       structure        & \textcolor{black}{expression for}         &  \textcolor{black}{expression for}          \\
  &calculations      & calculations      & $E^{0}( {\bf R^{3N}})$ & $E^{0}( {\bf R^{3N}})$   \\ 
  &   +                  &   +                    &          +             &         +                  \\
  & Path integral         & Classical              &  Path integral         & Classical  \\
Nuclei  &simulations            & Statistical            &  simulations           & Statistical \\
  &                 & Mechanics              &                 & Mechanics   \\
\\
 \\
\hline \hline
\end{tabular}
\end{table}
 
Teams C and D use analytical expressions for the PES.  These analytical expressions can be obtained in two completely different ways.
The analytical expressions can be obtained by fitting  ab-initio results obtained for water clusters and/or liquid configurations.
We shall denote this type of potentials as analytical ab-initio potentials. 
The second possibility is to propose an analytical expression 
for the potential with some free  parameters that can be chosen to reproduce some selected thermodynamic properties. 
We shall denote this second class as empirical potentials. 
Thermodynamic properties (i.e enthalpy, Gibbs free energy, ..) are functionals of U$({\bf R^{3N}})$. 
One could state that in analytical ab-initio potentials the parameters of the fit are determined to reproduce U $( {\bf R^{3N}}  )$ 
whereas in empirical potentials the parameters are determined to reproduce certain functionals of U$({\bf R^{3N}})$ ( i.e density, enthalpy, diffusion 
coefficients).

\textcolor{black}{ It is important when developing analytical potentials that "representative configurations" of the system 
are selected for the fit. By representative we mean configurations with a reasonable statistical occurrence (i.e
with a non-negligible value of their Boltzmann factor). Thus, the target, in principle, is not to reproduce 
the energy of any arbitrary configuration (including, for instance, configurations of very high 
energy where the water molecules overlap significantly), but rather properly describe  
the energy of those configurations of the ${\bf R^{3N}}$ space that have a reasonable probability of being found. 
Obviously the value of the potential parameters may depend on the configurations used for the fit either  explicitly as when using 
ab-initio inspired potentials, or implicitly as when using empirical potentials where the properties and 
selected thermodynamic states determine implicitly which configurations are entering in the fit. }

   When the description of the intermolecular energy is done with an analytical expression and  
nuclear quantum effects are used to describe the motion of the nuclei your approach is of type C.\cite{rossky85,ARPC_1986_37_0401_nolotengo,JCP_2006_125_054512,JCP_2007_127_074506,paesani09,vega10,Manolopoulos_2009,Carl_2009,ramirez14}
Simulations of analytical ab-initio potentials should be performed within the framework C, since when your PES was designed to reproduce ab-initio results
you should expect to reproduce water properties only when nuclear quantum effects are included. 
\textcolor{black}{ In team D, an analytical expression is used to describe the PES and the nuclear quantum 
effects are neglected (i.e it assumed that the motion of the nuclei can be described by classical statistical mechanics).} 
In the case of empirical potentials you could use approaches 
C or D. In fact you could determine the potential parameters  to reproduce experimental properties when nuclear quantum effects are included or you could determine 
the parameters of the potential to reproduce experimental properties within classical simulations. 
It is important to point out that if one has a good empirical  potential
model of water of type D (i.e one using classical statistical mechanics) and one tries  to use it within 
the formalism of type C (i.e including nuclear quantum effects) the model will not work. 
This is  because then, nuclear quantum effects will be counted twice, once through the fitting to
experimental properties and the other through the use of quantum simulations.\cite{Carl_2009,ramirez14} 
Group D, is by far the most popular. 
For this reason it is useful to classify the types of models that are often found within this family of potentials. 

\begin{enumerate}

\item{ \textcolor{black}{Ab initio potentials. } 
For these models, analytical expressions are used to reproduce either high level ab-initio results for small water clusters
(TTM2-F\cite{ttm2_f}, TTM3-F\cite{ttm3_f}, CCpol23\cite{szalewicz,szalewicz_2} ), or DFT results of condensed matter (Neural Network potentials\cite{neural_network_1} ) or both as for the MB-pol model\cite{paesani_model}.  Certainly quantum calculations are performed to develop these potentials.  However, instead of solving the Schr\"odinger equation 
on the fly to determine the energy of  each configuration (as you would do in teams A and B), here you assume that the fit used to reproduce the results of some water clusters and/or some
selected configurations, can be used for any configuration. Obviously, assuming that a good fit obtained for a small cluster water should also work in condensed 
matter is an approximation.  In the case
of Neural Network potentials\cite{neural_network_1,neural_network_2,neural_network_3} your results are obtained for 
condensed matter, but it is not clear if a neural network trained at a certain density and phase will also work for other densities
and/or phases.}

\item{ Empirical potentials. }
  The family of empirical potentials is large and several sub-classes could be identified.
  \textcolor{black}{Our classification of empirical models is based on the way electrostatic interactions are described.} 

\begin{enumerate}
\item{ Coarse grain models. The term coarse grained
is typically used for potential models that do not use partial charges in the description of the PES.\cite{review_coarse_grained}
Examples are the primitive model of water of Kolafa and Nezbeda\cite{kolafa_nezbeda_model_1,kolafa_nezbeda_model_2,nezbeda_review_pccp_2011} and
its modifications often used in combination with Wertheim's SAFT theory\cite{wertheim,chapman_jackson_gubbins,MP_1997_92_0135,JCP_1997_106_04168,FPE_2002_194_0087,lourdes_water,MP_2006_104_3561},  
the Mercedes Benz model\cite{mercedes_benz_water},  the
mW model.\cite{mw} and the ELBA model\cite{elba}. This last model does not have partial charges but incorporates
an ideal dipole moment on the oxygen atom. }

\item { Non-polarizable models. In these models a LJ center (or similar \cite{JPCB_1998_102_07470} ) is located
on the position of the oxygen atom. Models differ in the number and location 
of the partial charges. Three charges located on the atoms  (TIP3P, SPC), three charges with one charge
out of the atom positions (TIP4P family), four charges\cite{mahoney00} (TIP5P) or five charges\cite{nvde} (NvDE). In these models
the magnitude of the partial charges does not depend on the local environment. }

\item { Polarizable models. These are similar to the non-polarizable models, but now the partial
charges (or the molecular dipole moments) depend on the environment.\cite{paricaud05,review_polarizables,kolafa_polarizable,wang13,baranyai13,tavan14}. Typically polarizability is introduced either by allowing each molecule to develop an induced 
dipole moment in response to the local electric field or by \textcolor{black} { using the concept of charge transfer where 
part of the charge of one molecule (atom) is transferred to neighboring  molecules (atoms)\cite{charge_transfer_rick}. Strictly speaking polarizable models
are not analytical potentials (in a mathematical sense) as the energy must be obtained through an iterative process. However they are only one order of magnitude
more expensive (from a computational point of view) than non-polarizable models, in contrast to quantum calculations of type A or B which are about four orders of 
magnitude slower. For this reason we have included polarizable potentials in team D. Notice also that some of the ab initio potentials are polarizable.}}

\end{enumerate}

\end{enumerate}

\textcolor{black}{ Within each type of potentials described in group D  (i.e analytical potentials) 
 one could find two subsets, one in which the molecules are treated as rigid entities (rigid models) and those in which 
 flexibility is incorporated (flexible models). } 

\begin{figure}[!h]\centering
\includegraphics[clip,width=0.65\textwidth,angle=-0]{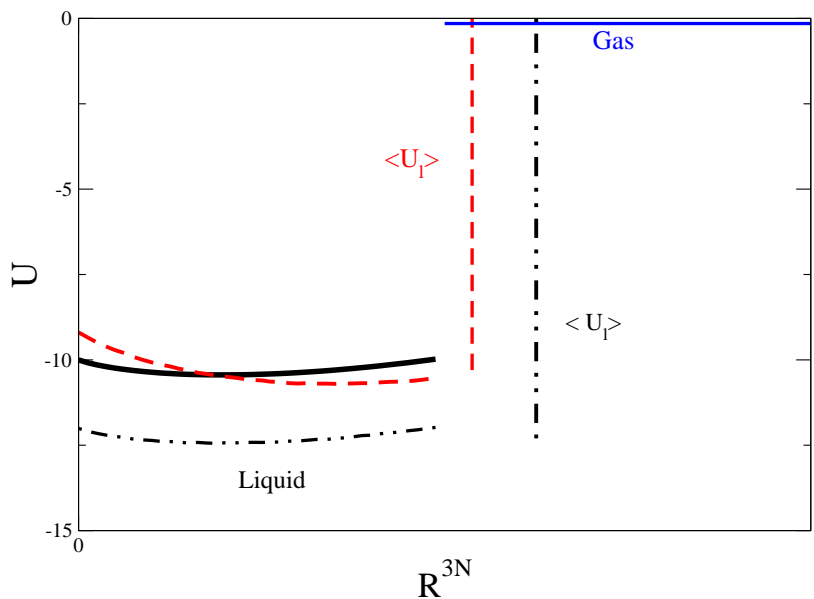}
\caption{ \textcolor{black}{Sketch of the PES surface as obtained from 
first principles for the liquid (configurations on the left hand side, black solid line) and vapor (configurations on the right hand 
 side, blue solid line ). 
 Certainly  $R^{3N}$ is multidimensional, so our presentation as an 1D object ( x axis) 
 is only an sketch. 
 Red dashed lines: sketch of the PES for a model reproducing the values of $<U_{l}>$ 
(the average residual internal energy of the liquid)  and the vaporization 
enthalpy $\Delta H_v$, but not the gradient of U, as TIP4P. 
Black dashed-dotted lines: Sketch of a model reproducing the gradient of U but neither $<U_{l}>$ nor $\Delta H_v$ as TIP4P/2005. 
 Vertical lines represent the value of $<U_{l}>$  which is one the main 
contributions to the vaporization enthalpy. }  }
\label{sketch_energia}
\end{figure}

 Non-polarizable models are by far the most used in the modelling of water. 
 For these models  U$({\bf R^{3N}}  )$ is usually described as:
\begin{equation}
\label{potencial_empirico}
 U ( {\bf R^{3N}}  ) \simeq    \sum_{i=1}^{\textcolor{black}{i=N-1}}   \sum_{\textcolor{black}{j=i+1}}^{j=N}
 4 \epsilon_{LJ} [ (\sigma/R_{ij})^{12} - (\sigma/R_{ij} )^{6} ]
+  \sum_{i=1}^{i=N-1}   \sum_{j=i+1}^{j=N} 
   \sum _{\alpha} \sum_{\beta}
  \frac{ q_{i_{\alpha}} q_{j_{\beta}}  } { 4 \pi \epsilon_{0} R_{i_{\alpha}j_{\beta}}  },
\end{equation}
where the indices
$\alpha$, $\beta$ run over the partial charges of each molecule, and 
there is only one LJ center per molecule (located on the oxygen atom).
 One of the main conclusions of the last twenty years, is that empirical non-polarizable potentials
such as SPC/E, TIP4P-Ew\cite{horn04} or TIP4P/2005\cite{abascal05b} that overestimate the 
vaporization enthalpy $\Delta H_{v}$  
provide a better description of water than those that try to reproduce it. 
The vaporization enthalpy can be estimated (when far from the critical point) as $\Delta H_{v} = -<U_{l}> + n R T$ where
$<U_{l}>$ is the residual internal energy of the liquid and n is the number of moles.
For this reason models as SPC/E, TIP4P-Ew, TIP4P/2005, underestimate $<U_{l}>$ and only get closer values 
to experiment for $\Delta H_{v}$ when  an "ad hoc" term, the polarization energy,  is added\cite{berendsen87,vega06}.
Why models that do not reproduce the vaporization enthalpy provide a better description of 
water properties ? A possible explanation is that these models try to reproduce the gradient of the intermolecular
energy (i.e the forces) rather than the absolute values of the intermolecular energy, providing an overall 
better description of the landscape for the liquid phase of the intermolecular energy.   
A graphical summary of this idea is presented  in Figure \ref{sketch_energia}. In the sketch of this 
figure it is qualitatively illustrated  how a potential may describe well U but not 
its gradient (as for instance  TIP4P) whereas another model may describe reasonably well the gradient but 
not U (as for instance TIP4P/2005). It is now clear that for non-polarizable models of water models it 
is not possible to simultaneously reproduce both U and the gradient of U, and that a better water model is obtained
when reproducing the gradient rather than the energy. 
 If your description of the gradient of U is reasonable, then the configurations generated 
 along the Markov chain of the MC run or along the trajectory in the MD run, would indeed be representative of those
 appearing in real water.  
\textcolor{black}{Having values of U in the liquid phase, shifted by a constant relative to the exact ones, 
will not influence the 
relative probabilities between two different configurations in this phase since it is proportional to the 
Boltzmann factor of their energy difference, and this difference remains unchanged if the energy of 
both configurations is shifted by a constant.  
The drawback is that this shift would be much smaller in the gas phase. Therefore,liquid-vapor coexistence properties 
(vapor pressures, vaporization enthalpies) will be affected reflecting that the relative probabilities between 
configurations of these two phases will not be described properly.}

 After presenting the different options (ABCD) in the modelling of water there is an interesting question: 
 Should one uses the approach  A, or B, or C or D when modelling water ?
 Obviously you should use the approach that is more convenient for the problem you have in mind. Therefore, there is
no unique answer to this question. 
Depending on the
problem it may be more convenient to use the approach A, B, C or D. For instance, it is difficult
to think how empirical potentials  can contribute to problems where water is involved in chemical reactions, 
or when computing electronic spectra.
At the same time, it is difficult that approaches A or B can attack problems
involving hundred of thousand of water molecules or very long times (for instance,
nucleation and supercooled water\cite{abascal10,liu12,bresme2014} or the conformational changes in proteins). 
Our point of view is that the four approaches are complementary. In fact 
it is becoming more common now to be at conferences about water where scientists of the four types of modelling 
are presenting their results.  These four approaches in the modelling of water will continue in the future. 

  Another different question is:  which 
approach provides an overall better description of water after ignoring chemical reactions and electronic spectra ? 
In principle results of approach A should be the only ones able to describe all the experimental
properties of water. Approach A is,  in principle, exact. However the reader may be surprised to learn that the 
results of approaches A and B are still far from describing the experimental properties of water. 
The reason is that we are not solving the Schr\"odinger
equation exactly. 
Reliable methods such as  MP2 or coupled cluster  become too expensive (except for small water clusters 
\cite{coupled_cluster_xantheas} )  
and they are not feasible right now for the system sizes and simulation times required to obtain the thermodynamic 
properties. Computationally cheaper methods such as DFT use approximate functionals.
Typically a set of letters is designed to describe the approximated functionals used in the 
calculations as for instance  PBE0, B3LYP.\cite{dft_review_1,dft_review_2,levine_book,jensen_book} 

  A possible way of tracking progress in the field is to perform an extensive
 comparison between calculated and experimental values for a number of selected properties. 
 We recently propose such a comparison and suggested a criterion  to obtain a numerical score.\cite{vega_perspective} 
 When agreement with experiment is good you obtain a high score. When agreement with experiment is low you obtain a 
low score.  Results for TIP4P/2005 are shown in Table \ref{score}. As it can be seen TIP4P/2005 got an score of $7.2$ points out of 10.  In our opinion obtaining
a higher score in the test means that you are describing better the PES of water.

\begin{table}
\caption{Scoring summary of the TIP4P/2005 (see Table IV of Ref. \cite{vega_perspective}  for further 
details) and surface (PES,DMS or both) required to determine the property. Strictly speaking for the dielectric
constant, the PS is also needed. However the contribution of this surface to the final value of the dielectric 
constant is small ( of about one per cent in condensed phases of water). $TMD$, $T_m$,$T_c$ stand for the
temperature of maximum density at room pressure, the normal melting temperature of ice Ih and the critical temperature
respectively. EOS refers to the equation of state.}
\label{score}
\begin{tabular}{lccccc}
\hline \hline
Property & TIP4P/2005 & Surface   \\
\hline
Enthalpy of phase change &  5.0 & PES \\
Critical point properties &  7.3 & PES \\
Surface tension &  9.0 &  PES \\
Melting properties &  8.8 & PES \\
Orthobaric densities and TMD &  8.5 & PES \\
Isothermal compressibility &  9.0 & PES \\
Gas properties & 0.0 & PES \\
Heat capacity at constant pressure &  3.5 &  PES\\
Static dielectric constant & 2.7 &  PES+DMS(+PS) \\
T$_m$-TMD-T$_c$ ratios &  8.3 &  PES\\
Densities of ice polymorphs &  8.8 & PES  \\
EOS at high pressure &  10 &  PES \\
Self diffusion coefficient  &  8.0 & PES  \\
Shear viscosity &  9.5 &  PES  \\
Orientational relaxation time  &  9.0 & PES   \\
Structure &  7.5 &  PES \\
Phase diagram  &  8.0 & PES  \\
 \\
Final score &  7.2 &    \\
\hline \hline
\end{tabular}
\end{table}

 The water test includes the comparison to experimental properties of the gas, liquid and solid phases of water.
Therefore the water test evaluates the capacity of the model to reproduce the PES under quite different conditions.
\textcolor{black}{ The PES depends formally on ${\bf R^{3N}}$, but for systems under periodic boundary conditions the volume of the
system V should also be provided. When performing a simulation 
at a certain value of N,V and T only configurations having a non negligible statistical weight will be found.
Let us denote this subset of configurations as  ${\bf R^{3N}}^{*}$. Obviously, the subset of explored configurations 
will be a function of the number density of the system   $d= ( N/V )$, the temperature, and in the case of solid
phases, the geometrical constrains imposed by the lattice $\Omega$. Therefore ${\bf R^{3N}}^{*}$ is a function 
of d,T and $\Omega$.  
Recently it has been shown how a polarizable model\cite{wang13} was able to obtain a higher 
score than a good non-polarizable one in the water test. 
 That makes sense and point out the existence of progress in the field. 
 The main reason for the higher score was an improvement in the score 
for those properties that depend on the description of the PES at low values of the density (properties of the gas, the virial coefficients, 
 vapor pressure, critical pressure) while keeping a good score for condensed matter properties.  Non-polarizable models, are designed to describe the condensed matter phases but are unable to describe the properties of the gas.} 

  In this section we have presented different possible approaches in the modelling of water. The central idea of this 
 paper is related to way the dielectric constant should be computed, when modelling water within the approaches 
 C and/or D. 
 In Section IV we will describe how the dielectric constant is commonly obtained in computer simulations and 
 in experiments.  But before, and to illustrate the reasons behind the main point of this paper, 
 it seems pertinent to summarize some basic ideas of quantum chemistry. 
 In particular how the energy of a system can be obtained from quantum calculations, both in the absence and in 
 the presence of an electric field.

 \section{  A little bit of Quantum Chemistry :  potential energy and dipole moment surfaces.  }

We shall start by presenting two of the most important surfaces in the modelling of water, the 
potential energy surface (PES)  and the dipole moment surface (DMS). We shall first explain how they are obtained 
from quantum calculations, and secondly we will discuss which properties are determined by the PES and which ones by the DMS. 

  Within the Born-Oppenheimer approximation one should solve the Schr\"odinger equation for a certain
fixed configuration of the nuclei of the system ${\bf R^{3N}}= {\bf R_{1}, R_{2}, ... R_{3N}}$. 
The positions of the  $n_{e}$
electrons are denoted as ${\bf \tau^{n_{e}}} = {\bf \tau_{1},\tau_{2}, ... \tau_{n_{e}}}$ 
(obviously for water $n_e=10N$), 
where ${\bf \tau_{i}}$ stands 
for the coordinates of position and spin of electron i (i.e ${\bf r_{i} } s_{i}$ ) .
In the absence of an electric 
field ( $E_{el}$ )  the energy of the system can be obtained by solving the Schr\"odinger equation:

\begin{equation}
\label{schrodinger_referencia}
  \hat{H}^{0} \Psi^{0} ({\bf \tau^{n_{e}} ; R^{3N} } )  = E^{0}( {\bf R^{3N} })   \Psi^{0}({\bf \tau^{n_{e}}; R^{3N} } )   
\end{equation}

The superscript $^{0}$ indicates the absence of an electric field. The hat indicates an operator. 
Unless other thing is stated we shall focus on the ground state, so that the energy and wave function refer to that of the ground state.
Notice that $\hat{H}^{0}$ includes the internuclear Coulombic repulsion energy. 
The total dipole moment of the system ${\bf M^{0}}$ is obtained as\cite{levine_book,jensen_book}:

\begin{equation}
\label{momento_dipolar_molecula}
 {\bf M^{0} } ( {\bf R^{3N}} )  = e ( \sum_{\gamma } Z_{\gamma } {\bf {R}_{\gamma} } - \int {\bf r} \rho^{0}( {\bf r} ) d {\bf r } ) 
\end{equation}

where $e$ is the magnitude of the electron charge, $Z_{\gamma}$ is the
atomic number of atom $\gamma$
and $\rho^{0}$ is the electron density at  point ${\bf r}$, 
 which can be easily obtained from the wave function as \cite{levine_book}:

\begin{equation}
\label{rho_r}
\rho^{0} ( {\bf r} ) = n_{e}   \int...\int  \Psi^{*}({\bf r },s_{1},{\bf \tau_2},...{\bf \tau_{n_{e}} } ) 
 \Psi({\bf r },s_{1},{\bf \tau_2},...{\bf \tau_{n_{e}} } ) 
 ds_{1}   d{\bf \tau_2} ... d{\bf \tau_{n_{e}}}  
\end{equation}

 Notice that both the energy and the dipole moment of the system depend on the positions 
of the nuclei, therefore they are functions of  ${\bf R^{3N}}$. Determining $E^{0}$  for different configurations of the 
nuclei provides the potential energy surface (PES), $E^{0}( {\bf R^{3N} })$ . Determining the dipole moment for different positions of the nuclei 
provides the dipole moment surface (DMS),  $ {\bf M^{0} } ( {\bf R^{3N}} )$ . 
The existence of two different surfaces when describing properties of a system is well known in the
quantum chemistry community\cite{pes_monomer_famosa,two_surfaces_in_water_clusters_ab_initio} but probably less well known 
in the community performing condensed matter simulations with empirical potentials.

The energy $E^{0}$  and the  dipole moment ${\bf M^{0}}$  are observables so that in principle they can be measured.
There is an operator for each of these two magnitudes, and it is easy to determine their values once the
wave function is known.  However the dipole moment of each individual molecule (in a certain ${\bf R^{3N}}$ configuration) 
can not be measured experimentally and there is no operator 
linked to the dipole moment of a single molecule in condensed matter. 
\textcolor{black} {The same is also true for the total quadrupole moment 
of a system. It can also be determined experimentally by using an inhomogeneous 
electric field. However it is  not possible to determine the quadrupole moment of each individual molecule (in a certain ${\bf R^{3N}}$ configuration) 
and there is no operator linked to the quadrupole moment of a molecule in condensed matter. } 
The problem when determining the molecular
dipole/quadrupole moment of a molecule in condensed matter is that for each point of the space ${\bf r}$, with an electronic
density $\rho^{0}( {\bf r } ) $, one must decide somewhat arbitrarily to which molecule of the system this point {\bf r} belongs.
There is no a unique way of doing that and for this reason there is no a unique way of determining the 
dipole moment of a molecule in condensed matter\cite{jensen_book,leach_book,bader_book}. 
The dipole moment of an individual molecule is not needed either to compute the energy of a certain configuration 
or to compute the total dipole moment of the system in a certain configuration. However it may be useful to rationalize the 
obtained results. 
Defining the dipole moment of a molecule in condensed matter
is useful as a pedagogical concept, as it allows one to better understand the properties of condensed matter.
In the same way the partial charge of an atom in a molecule can not be measured. In fact
there is no operator to determine partial charges. Partial charges are only useful to obtain a graphical
simple picture of the charge distribution within  the molecule or eventually to obtain an initial educated
first trial in the design of empirical potentials.
\textcolor{black}{ Although partial charges can not be determined  in a unique way it is certainly possible to conceive that
  a certain prescription yields partial charges that can be used with success in the development of a force
  field for a given molecule.} 

 Let us now apply a uniform static electric field ${\bf E_{el}}$.
Let us assume that the electric field  is applied along the z direction and its modulus is $E_{el}$.
The energy of the system for a certain configuration of the nuclei  ${\bf R^{3N}}$ is obtained
by solving the Schr\"odinger equation:

\begin{equation}
\label{schorodinger_con_field}
  ( \hat{H}^{0} - E_{el}  \hat{M}_{z} ) \Psi ({\bf \tau^{n_{e}}} ;E_{el}, {\bf R^{3N}} )  = E( {\bf R^{3N}},E_{el} ) 
\Psi ({\bf \tau^{n_{e}}} ;E_{el}, {\bf R^{3N}} ) 
\end{equation}

The total dipole moment of the system $M$ is obtained as:

\begin{equation}
\label{momento_dipolar_molecula_con_field}
 {\bf M} ( {\bf R^{3N}} )  = e ( \sum_{\gamma } Z_{\gamma } {\bf {R}_{\gamma} } - \int {\bf r} \rho( {\bf r} ) d {\bf r }  ) 
\end{equation}

where $\rho$ (without any subscript) is the electron density in the presence of the field, which can be obtained easily from the wave function.
 It follows from Eq. \ref{schorodinger_con_field} that the energy of the system in the presence of the external field can be written as: 
\begin{equation}
\label{desglose}
E( {\bf R^{3N}},E_{el} ) =  \int  \Psi^{*}  \hat{H}^{0}  \Psi  d{\bf \tau^{n_{e}} }  -  E_{el}  \;  M_z 
\end{equation}

According to this the energy can be divided into two contributions. The first one is 
the intermolecular energy, and the second one is the contribution due to the interaction of the system 
with the external field. Notice however, that even the first term depends on the external electric field since 
the wave function $\Psi$ depends on the external field and it is not identical to $\Psi^{0}$. 
If the external field $E_{el}$ is weak,  one can use quantum perturbation theory 
using the external field as the coupling parameter to estimate the energy of 
the system.  In that case (to second order in  $E_{el}$ ) one obtains:

\begin{equation}
\label{perturbation}
E( {\bf R^{3N}},E_{el} ) = E^{0} -  E_{el} \;\;   M^{0}_{z} + E_{el}^2 \; \sum_{j} 
 \frac{ | \int  (\Psi^{0})^{*}  \hat{M}_{z}  \Psi^{0}_{j} d{\bf \tau^{n_{e}}}  |^2 } 
{( E^{0} - E^{0}_{j} ) } + .... 
 \end{equation}
where the subindex j, labels the excited states of the system in the absence of the external field.
 The previous equation can be written as:

\begin{equation}
\label{perturbation_2}
E ( {\bf R^{3N}},E_{el} ) = E^{0} -  E_{el} \;\;   M^{0}_{z} - \frac{1}{2} E_{el}^2 \;  \alpha_{zz}^{0} + ... 
 \end{equation}
where $\alpha_{zz}$ is the zz component of the polarizability tensor. 
It follows that:

\begin{equation}
\label{perturbation_3}
   M^{0}_{z} ( {\bf R^{3N}} )  =  - \left( \frac{ dE( {\bf R^{3N}},E_{el} )   } { dE_{el}} \right)_{\tiny E_{el}=0} 
 \end{equation}

\begin{equation}
\label{perturbation_5}
  \alpha^{0}_{zz} ({\bf R^{3N}} ) =   \left( \frac{ dM_{z}( {\bf R^{3N}},E_{el} )   } { dE_{el}} \right)_{\tiny E_{el}=0} =   - \left( \frac{ d^{2}E( {\bf R^{3N}},E_{el} )   } { dE_{el}^{2}} \right)_{\tiny E_{el}=0}  
 \end{equation}

 The energy of the system in the absence of the external field $E^{0}$ defines the potential energy surface.
The first derivative of the energy with respect to the external field ( at zero external field) defines the dipole 
moment surface DMS ( $M_{z}^{0}$ ). Strictly speaking the dipole moment surface is formed by three different 
surfaces ( i.e $M_{x}^{0}$, $M_{y}^{0}$, $M_{z}^{0}$ ). 
The second derivative of the energy with respect to the external field is the 
polarizability surface PS\cite{nh3_ps_surface}. Obviously the PS is formed by nine components and  is a tensor. Each component represents 
a different second derivative (xx, xy, ... zz). For this reason the PS is formed by 9 different surfaces. 
Notice that polarizability is related to the derivative of the polarization of the system 
with respect to the external field.  After introducing the PES and DMS surfaces, it is interesting to raise the following question: which properties are obtained from the PES and which ones from the DMS?

%
%
  In Table \ref{score}, a list of the properties that can be obtained once the PES is known is presented. 
As  can be seen  the knowledge of the PES is enough to compute practically all experimental properties of 
the system.  In fact to perform Monte Carlo simulations, one only needs to know the energy of each configuration (and its
gradient too in the case of Molecular Dynamics). 
The only property that can not be evaluated, even after the PES is known, is the dielectric constant. To determine the 
dielectric constant both the PES and the DMS are needed (and also the polarizability surface PS although the contribution of this
surface in the case of water is rather small).
In the absence of the electric field all properties of water can be obtained from the PES. In this case 
you should not care at all about the DMS and PS surfaces  because without the presence of the electric 
field they play no role!
In the physics of water (or in that of any other substance or system) the dielectric 
constant is a property that matters only when applying an electric field to the sample. 
Due to this particularity, it is interesting to discuss in some detail the procedure used to 
determine the dielectric constant both in experiments and in computer simulations.

\section{ The dielectric constant }
In experiments the dielectric constant is obtained from the relation  between  
the polarization $<P_{z}>$  and  the electric field:

\begin{equation}
\label{polarizacion}
   <P_{z}> =   \frac{ <M_{z}> }{V} = \chi  E_{el} = \epsilon_{0} ( \epsilon_{r} -1 ) E_{el} 
\end{equation}
 where $\chi$ is the susceptibility, $\epsilon_{0}$ is the permittivity of vacuum and $\epsilon_{r} = \epsilon / \epsilon_{0}$ ( the ratio of the 
permittivity of the medium with respect to vacuum)  is the dielectric constant. 
 In general the electric field acting on the sample, $E_{el}$, is  not identical to the applied 
external field $E_{ext}$, as surface charges are formed at the interfaces 
of the sample, and these surface charges generated an additional
contribution to the field.\cite{neumann83,kolafa_dielectric_constant}  However, if the sample is confined within a conductor (i.e the dielectric around
the sample has an infinite dielectric constant) then $E_{el}$ becomes identical to $E_{ext}$. For simplicity
we shall assume that this is the setup used both in experiments and in the calculations so that $E_{el}$ and $E_{ext}$
are identical (i.e we are using conducting boundary conditions).  For weak electric fields the relation between $<P_{z}>$ and  $E_{el}$ is linear and the slope defines the value of the dielectric constant. Therefore:
 
\begin{equation}
\label{dielectric_version_1}
 \epsilon_{r}  = 1 + \frac{1}{\epsilon_{0}}  \left( \frac{ d<P_{z}> } { dE_{el}} \right)_{\tiny E_{el}=0}  = 
 1 + \frac{1}{\epsilon_{0} \; V}   \left( \frac{d<M_{z}>} {dE_{el}}  \right)_{\tiny E_{el}=0} 
\end{equation}

 Let us now assume that the motion of the nuclei can be described using classical statistical mechanics  
(the formalism can be easily extended to the case where one incorporates nuclear quantum effects).  
Then (in the NVT ensemble):

\begin{equation}
\label{polarizacion_NVT}
 < M_{z} >  
 = \frac{ \int {    M_{z}( {\bf R^{3N}},E_{el} )    e^{- \beta E({\bf R^{3N}},E_{el})}   d{\bf R^{3N}} }  } 
        { \int {                                      e^{- \beta E({\bf R^{3N}},E_{el})}   d{\bf R^{3N}} }  } 
\end{equation}
 
Notice that both $M_{z}( {\bf R^{3N}},E_{el}) $  and $E({\bf R^{3N}},E_{el}) $ are functions of the position of the nuclei and of the electric field. If the zero of energies were chosen as the energy  of N isolated water molecules 
in the absence of the field then this change, of course, would not affect the value of $< M_{z} >$.
To evaluate $\epsilon_{r}$ all that is needed is to evaluate the derivative of $<M_z>$ with respect to $E_{el}$ at zero external field (see Eq(\ref{dielectric_version_1}).
By using the expression obtained to first order from quantum perturbation theory for $E({\bf R^{3N}},E_{el}) $ one
obtains:

\begin{equation}
\label{dielectric_version_2}
\epsilon_{r} = 1 + \frac{1}{\epsilon_{0} \; V} \left<  \left(\frac{ dM_{z}( {\bf R^{3N}},E_{el} )     } {dE_{el}} \right)_{\tiny E_{el}=0} \right>_{0} + \frac{\beta}{\epsilon_{0} V} (  <  ( M^{0}_{z} )^{2} >_{0}  - <  M^{0}_{z} >^{2}_{0}    )    
\end{equation}

 where the $< X >_{0}$ represents the canonical average of property X over configurations generated in the 
absence of the electric field. 
Although the discussion can be formulated for a general case, for simplicity let us focus on an isotropic 
phase (for instance a liquid phase). In this case the value of $<M^{0}_{z}>_{0}$ is zero (there is no net polarization in the absence of the field), and the directions x, y, z are equivalent so that better statistics is obtained by averaging the results over the three axis. The final expression is: 

\begin{equation}
  \epsilon_{r} = 1 + \frac{1}{\epsilon_{0} V} \left< \left(\frac{dM_{z}}{dE}\right)_{\tiny E_{el}=0} \right>_{0} +   \frac{  \rho  } { \epsilon_{0} 3 k T }
  \frac{ < ( {\bf M^{0}} )^2 >_{0} } {N}  = \epsilon_{r,\infty} + \frac{ \rho  } { \epsilon_{0}  3 k T } 
  \frac{ < ( {\bf M^{0}} )^2 >_{0} } {N}  
 \label{eq_fluctuaciones}
\end{equation}
This is the expression in the SI system of units. To obtain the corresponding formula in the CGS (often used in simulations) one should replace $\epsilon_{0}$ by $1/(4\pi)$ 
in the previous expression. 
The dielectric constant is the sum of three  contributions. The first one is a constant with value one. 
The second contribution accounts for the average change 
of the polarization of the system for an instantaneous configuration when an external field is applied.
The sum of these two terms is usually denoted as  $\epsilon_{r,\infty}$. 
The third  contribution accounts for the polarization induced in the system by the alignment of the permanent 
dipole moments of the molecules with the external electric field. 
Let us briefly comment on the value of $\epsilon_{r,\infty}$. It can be determined from experiments by using 
an electric field of high frequency. In fact when the electric field has a high frequency, the permanent 
dipole moment of the molecules of water are unable to align with the external field within the time scale
of one oscillation. For this reason it is possible to determine $\epsilon_{r,\infty}$ from experiments by 
using high frequency electric fields. It can also be determined from theoretical calculations. The value 
of $\epsilon_{r,\infty}$ for water is of about 1.8 both for pure water and 
for ice Ih \cite{johari_1976,galli_infinita}.
Since the dielectric constants of liquid water and ice Ih at the melting point are 88 and 94 respectively
it is clear that, in condensed matter, the largest contribution to the dielectric constant comes by far, from the last  
term on the right hand side of Eq.\ref{eq_fluctuaciones}. The dielectric constant of water is high, not because the external field 
significantly changes the polarization of individual configurations, but because it significantly changes the probability of 
each individual configuration in the ensemble by increasing the probability of configurations with large polarization.

\textcolor{black}{  The way to compute $\epsilon_{r}$ in computer simulation is rather straightforward. One performs simulations 
in the absence of the electric field. One only needs the PES to perform those simulations. You store in the hard
disk, say, 10000 independent configurations for later  analysis. For each configuration one evaluates 
its dipole moment ${\bf M^0}$ (which is obtained from the DMS) and the derivative of $M_z$ with respect to the external field 
evaluated at zero external field (which is obtained from the PS). Obviously expressions for the DMS and PS are needed.} 
After obtaining the average over the 10000 configurations one obtains  the value of the
dielectric constant. In summary one only needs the PES to generate the trajectory over the phase space, and then 
for the analysis leading to the dielectric constant one also needs the DMS and PS surfaces. 

  Now we will present the main point of this paper.

\section{ One molecule,  two surfaces. }

 The PES and the DMS are two functions that depend on ${\bf R^{3N}}$. They are two surfaces on the imaginary plane where
${\bf R^{3N}}$ are the independent variables. Both PES and DMS can be obtained from the wave function.

\subsection{ One side of the mistake:  transferring from the PES to the DMS }
 
Empirical potentials are simple expressions designed to describe (although in an approximate way) the PES. 
They usually contain parameters for the LJ part of the potential, and parameters (i.e partial charges) to describe
Coulombic like interactions.  

 Now it is time to introduce  the "dogma" that has been used implicitly by a number of
 people (including the author of this paper).\cite{jorgensen05,guillot02,vega_perspective} 
The "dogma" states  that "the partial charges " used to describe empirically the PES should also be used to describe empirically
the DMS.  According to the "dogma", it should be done in this way, and it would not be legal, possible or correct 
to do something different. 

 But ... if the PES and the DMS are two surfaces, why should we use the same set of fitting parameters to describe two different functions?
Let us assume that both the PES and the DMS are known from ab initio calculations. In the case we are using partial charges 
to describe empirically the PES and/or  DMS, one would expect that the parameters providing the best fit (i.e with 
the minimum of the average square deviation) for the PES would, in general, be different from those obtained to reproduce the 
DMS.  Therefore there is no conceptual reason why one could not use a different set of partial 
charges to describe the PES and the DMS (in contrast with the "dogma" that states that they should be identical).
  The main point of this paper is to point out that the implicit assumption that one should use the same partial 
 charges to describe the PES and the DMS is a "conceptual" mistake.  
Let us analyze whether leaving the "dogma" presents technical difficulties. 
 When performing simulations using an  empirical PES 
 one  stores a set of configurations on  the hard disk.  It is clear that now 
 you  can use whatever expression you want to obtain the dipole moment of the stored configurations.  
 There is no technical difficulty in doing that. One can write a program to generate the configurations 
 from a certain PES, and another one reading these configurations and obtaining  the DMS  using a different 
 set of parameters. In fact one does not need two programs. One could do that within one program. It is enough 
 to have two subroutines, one for the PES (which enters in the Markov chain or when computing 
 forces) and another one for the DMS (which enters to compute the dipole moment of each configuration). In the 
 case the PS is also considered, then another subroutine for the PS is needed. 
 Of course current popular programs (Gromacs\cite{hess08}, DLPOLY\cite{DLPOLY2}, Lammps\cite{lammps_program}...) do not allow one to do that because 
 they have been written respecting the "dogma". However, modified versions of these 
 codes leaving the dogma can be easily written. 

    In the design of empirical potentials for water we probably misunderstood the role of the dielectric constant.
  The dielectric constant it is not the property to look at to obtain a good PES .  It depends on two surfaces and 
  when one  fails in describing $\epsilon_{r}$ one does not know whether this is due to a good PES combined with a bad 
  DMS, to a bad PES combined to a good DMS, or to the combination of a bad PES and a bad DMS (although in this
  last case there is the possibility that one describes quite well the experimental value if the errors in 
the two surfaces cancel out partially). 
  The way to go suggested in this work is as follows. One first tries to develop an empirical expression for the PES, by reproducing
  as many experimental properties as possible (but eliminating the dielectric constant from the test). Once 
  you have a good PES, then you fit your empirical expression for the DMS by fitting to the experimental values 
  of the dielectric constant. 

  Once one  leaves the "dogma" there are many possibilities. 
For instance, one could use a model like TIP4P/2005 for the PES and 
use quantum chemistry, or a polarizable model  to determine ${\bf M^{0}}$  and the polarizability for the configurations stored on the hard disk.
\textcolor{black}{ In fact such approach has been used recently by Hamm to determine the 2D Raman THz spectra of water\cite{hamm_preprint,hamm_experiments} and
make a comparison with the experimental results. Also Skinner and coworkers
\cite{skinner_2014_dielectric}  found that it was possible to describe 
the low frequency region of the IR spectrum of water and ice Ih by using a non-polarizable model for the PES and a polarizable 
model for the DMS.}   
In the future it may be very interesting to determine  ${\bf M^{0}}$ from first principles for the configurations 
obtained by using an empirical potential. There are some lessons to be learnt from that. 
Probably  we have not fully appreciated the fact that the PES and the DMS are two different surfaces 
and there is no reason why both of them should be described by the same set of charges, parameters or methodologies.

As far as we know the dogma was challenged in at least three recent papers. In our previous work we used the "charge scaling" method (see discussion about this method below) for the DMS.\cite{aragones11a,aragones11b} 
The group of Skinner has also presented recently an example of "departure from the dogma". 
Skinner and co-workers developed the E3B model\cite{skinner_e3b},
a model that adds three body forces to a TIP4P like model. 
The addition of three body forces in principle should improve the description of the PES. However it was 
found that the E3B model did not improve the description
of the dielectric constant of water.  Why? Because once again 
the same set of charges were used to describe the PES and the DMS. 
However quite recently Skinner and co-workers, used 
the E3B model for the PES and a polarizable model to describe the DMS
with reasonable agreement with experimental results.\cite{skinner_2014_dielectric}  
Probably these two works can be regarded as the first excursions away from the "dogma". 
The idea is also "in the air" in the recent papers of Leontyev and Stuchebrukhov\cite{rusos1,rusos2,rusos3} where they 
suggested that the charges to be used in the PES of a non-polarizable 
model correspond to the scaled charges of a polarizable model ( assuming that they mean that the charges of the non-polarizable
model are used to obtain the PES and the charges of the polarizable model are for the DMS).  
We do hope that many more examples like these (i.e leaving the dogma) will come.

If one solves the Schr\"odinger equation 
exactly (as nature does), then from the exact wave function one obtains  both the exact PES and DMS. 
The power of  approaches A and B, is that as one gets a better and better wave function (or functional) one  will be able to obtain 
from the wave function (or from the electronic density) both an accurate PES and DMS. 
The assumption  that a simple empirical potential 
is able to describe all features of the PES is somewhat optimistic although one must admit that it is amazing 
how much can be described by such a simple approach. 
However, even admitting that an empirical potential with partial charges can do a reasonable job in describing the PES, assuming that 
the same partial charges are good to describe the DMS,  is simply "too much".
It is interesting to point out that the collaboration between teams A/B and D could be very useful to obtain accurate 
values for $\epsilon_{r}$. 

 We have described above how it is possible and simple to determine $\epsilon_{r}$ without invoking the "dogma" from 
 the expressions obtained from linear response theory. The dielectric constant can also be obtained by applying 
 a weak electric field. Once again, for simplicity we shall assume that the field acts on the z-axis and shall 
 use conducting boundary conditions. Then one has \cite{kolafa_dielectric_constant,aragones11b}:

\begin{equation}
\epsilon_{r} =  1 + \frac{<M_z>}{\epsilon_{0} E_{el}  V } = 1 + \frac{1}{\epsilon_{0}  E_{el}  V} 
\frac{ \int    M_z  e^{- \beta E({\bf R^{3N}},E_{el})}    d{\bf R^{3N}} }
     { \int         e^{- \beta E({\bf R^{3N}},E_{el})}    d{\bf R^{3N}} }  
\end{equation}

 For a weak electric field one can use first order perturbation theory both for $E({\bf R^{3N}},E_{el})$ and for $M_z$: 

\begin{equation}
\epsilon_{r} =  1 + \frac{1}{\epsilon_{0}  E_{el}  V}  \left<   M^0_{z} +  \left( \frac{dM_z}{dE_{el}} \right)_{\tiny E_{el}=0} \;\;\; E_{el} \right>_{\tiny E^{0} -  E_{el} \;   M^{0}_{z} }
\end{equation}

\textcolor{black}{  To evaluate this expression one needs to store on the hard disk configurations generated according to the Boltzmann 
distribution of $E^{0} -  E_{el} \;  M^{0}_{z}$ (so that the PES and DMS are needed). 
Once these configurations are saved, you simply evaluate 
the average of the value in the bracket (which requires to know both the DMS and the PS). 
Many standard MD and MC programs allow one to apply an external electric field.
The codes were written to obey the "dogma", so the same partial charges and/or multipoles are used for the PES and the DMS.
It is generally stated that for non-polarizable models the PS is zero. In the case of polarizable models the
PS is described by a simple electrostatic model describing how the DMS changes with the electric field.  
However these codes could  be easily modified to deviate from the "dogma", by simply  allowing different treatments when 
describing the PES, DMS and PS surfaces. As discussed previously the contribution of the PS to $\epsilon_r$ for water
at room temperature and pressure is small (of about 1 \%) so that error introduced by neglecting this contribution is small.} 

  We shall now illustrate a very simple example where we abandon the dogma. Although more complex 
 treatments could obtain much better results, the "$\lambda$" scaling is probably the simplest example to illustrate 
 the ideas of this paper at work.

\subsection{ The $\lambda$ scaling }
Let us assume that to describe the PES  one is using, in addition 
to the traditional LJ parameters, a set of partial charges. We shall denote the partial charges used to describe the 
PES as $q_{PES}$. Let us now assume that to describe the DMS one is using a set of charges that are identical to those 
used to describe the PES (and located at  the same positions) but scaled by a factor $\lambda$. Then it follows that:
 \begin{equation}
\label{charge_scaling}
    q_{DMS} = \lambda q_{PES}
 \end{equation}
 We shall denote with subscript $\lambda$ the properties that will follow when using the scaled charges for 
 the DMS (while using the original charges for the PES) and by $PES$ the 
 properties that will follow when using the same charges for the PES and the DMS. 
 It follows that:
 \begin{equation}
   {\bf M^{0}_{DMS}}  = \lambda {\bf M^{0}_{PES}}
 \end{equation}
 Implementing the ideas described above (and assuming for simplicity that the PS contribution is zero) one obtains:

\begin{equation}
  \epsilon_{r,\lambda} =  1  +  \frac{  \rho  \lambda^{2} } {\epsilon_{0} 3 k T }  \frac{ < ({\bf M^{0}_{PES}}) ^2 > } {N}  
 \label{eq_fluctuaciones_escalada}
\end{equation}

\begin{equation}
\label{epsilon_scaling}
 \epsilon_{r,\lambda} = 1 +   ( \epsilon_{r,PES} - 1) \lambda^2
\end{equation}
  Where we denote $\epsilon_{r,PES}$ as the value that will follow from evaluating the dielectric constant in the 
traditional way (i.e using the same partial charges for the PES and for the DMS). Several previous studies
suggested that the dipole moment of water in condensed matter's is of about 2.66 D \cite{saykally_clusters,JCP_2000_112_09206,gubskaya02,batista98,batista99,morrone_car,robertocar_ice,rick11_cojounudo_ctes_dielectricas_hielos}
.  In the TIP4P/2005 model all molecules have a dipole moment of 2.305 D . Then the value of 
$\lambda$ that follows from this reasoning is $\lambda=(2.66/2.305)=1.15$. 
Let us now evaluate the dielectric constant of water using this scaling. 

\begin{figure}
\label{epsilon_escalada}
 \includegraphics[clip,height=180pt,width=200pt,angle=0]{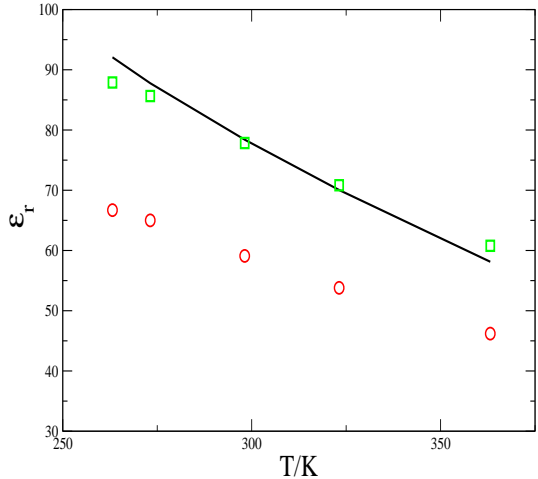}
\caption{ Dielectric constant of liquid water at room pressure as a function of the temperature. Line: experimental results.
 Circles: Results for TIP4P/2005 from Ref. \cite{kolafa_dielectric_constant}, and
       from the $\lambda$ scaling (with $\lambda=1.15$) as obtained in this work (squares). }
\end{figure}

The  results obtained are presented in Figure 2.  The 
dielectric constant of liquid water for TIP4P/2005 was taken from the 
recent work by Kolafa and Viererblova.\cite{kolafa_dielectric_constant} 
As can be seen the description of the dielectric constant of water is now
much better. At 298K the predicted value of $\epsilon_{r,\lambda}$ is 77.8 which should be compared to 78.5 which 
is the experimental value. Also the variation of the dielectric constant with temperature is now in better 
agreement with the experimental results.\cite{marivi_epsilon} With respect to ice Ih, the value of $\lambda$ required to bring the
simulation results of TIP4P/2005 into agreement with the experiments is  $\lambda=1.41$ which implies that
the dipole moment of the molecules of water in ice is about 3.25 D, in reasonable agreement with 
previous estimates  \cite{JCP_2000_112_09206,gubskaya02,batista98,batista99,morrone_car,robertocar_ice,rick11_cojounudo_ctes_dielectricas_hielos}. It is not possible to reproduce simultaneously the dielectric constant of liquid water and ice Ih using 
an  unique value of $\lambda$. The use of the $\lambda$ scaling would modify the score of TIP4P/2005 model in the 
block of dielectric properties (of course it will not affect the score in the rest of the properties). 
If an  unique value of $\lambda$ is used for the fluid and ice (i.e $\lambda=1.15$ ) then the score for the three properties
of the dielectric constant block would be 10(liquid), 3 (Ih) and 3 (ratio of the dielectric constant of ice and water) 
so  the average of this block would be 5.3. Using two different
values for $\lambda$ ( one for ice and another one for liquid) would dramatically increase the score of this section
since basically one would now reproduce the experimental results. Thus the use of the $\lambda$ scaling will 
increase the global score of TIP4P/2005 from $7.2$ to $7.4$ (when using the same value of $\lambda$ for all phases)
and to $7.6$ when using a different value of $\lambda$ for the liquid water and for ice. 
It is probably true that all previous discussions\cite{guillot02,jorgensen05,vega_perspective} 
about the ability of water models to describe the dielectric constant
should be revised, since all the reported values were obtained under the implicit assumption that 
the charges used to describe the PES should also be used to describe the DMS. 
Recently  the TIP4P model has been modified to obtain a "special purpose" model that improves the
description of the dielectric constant\cite{alejandre14} of water (while still using the "dogma"). Further work is needed 
to analyze if the improvement\cite{alejandre14} in the description of $\epsilon_{r}$ is  at the cost of deteriorating somewhat 
the PES, leading to an overall lower score in the water test (as compared to TIP4P/2005). 
Polarizable models, in principle, should improve the description of the dielectric constant of water, although
further work is needed to analyze if the improvement obtained for the liquid phase does also extend to the 
solid phases\cite{rick11_cojounudo_ctes_dielectricas_hielos}. Notice also that the use of the "dogma" is also
present in polarizable models, since  the charges/multipoles used to describe the PES are also used to describe 
the DMS.

\textcolor{black}{ Even in the case of polarizable models  leaving the "dogma" may result in  an improved description of
the dielectric properties of water (so  the main point of this work does not only apply to non-polarizable
models). In any case it seems  that when using polarizable models (especially those using  diffusive partial
Gaussian charges\cite{paricaud05,baranyai13}  rather than point like partial charges) the differences between the optimum set of charges needed
to reproduce the PES and those needed to reproduce the DMS are smaller than when using non-polarizable models.
Thus with polarizable models the need to use different charges for the PES and the DMS is reduced considerably.
However the option of  using  different approaches to describe both surfaces is still possible 
and the benefits of such a treatment remains to be explored.
Obviously in a quantum treatment, the same electron density should be
used to compute both the DMS and the PES (in fact in DFT the energy is obtained once the electron density is known).
However, an empirical  polarizable model is not identical to a quantum treatment so the option of using different approaches
for the PES and DMS could still be beneficial. Further work on this issue is needed before establishing definite conclusions.}

 The failure of all non-polarizable models in describing the dielectric constant of ice Ih was the "smoking gun",
announcing that something was totally wrong in our treatment of dielectric properties. 
The dielectric constant of ice Ih was not computed often for water 
models and that may explain our delay in understanding the situation. 
TIP4P/2005 was successful in describing  many properties of water indicating that it has a reasonable PES. 
The fact that dielectric constant of both  ice Ih and water was incorrect, but 
always much lower than the experimental value, was a clear indication that there should be a reason for that. In our 2011 paper 
we  indicated that this was a failure of the model, and that it was a consequence of the fact that the model is non-polarizable. 
In this work we go one step further. Our point is that there was something wrong but "in our mind". 
Lennard Jones centers, partial charges, polarizable models that respond to a local
 electric field, are just approximations aimed to describe the PES, which of course can only 
 be obtained from quantum mechanics. Forcing these entities to reproduce simultaneously two surfaces 
obtained from the quantum treatment (PES and DMS) was in, retrospective, a naive hope.

\subsection{ The other side of the mistake: transferring from the DMS to the PES.}

Let us consider a diluted solution of NaCl in water in the absence of an electric  field. The dipole moment of a certain 
configuration is given by Eq.\ref{momento_dipolar_molecula}.
In Eq.\ref{momento_dipolar_molecula} the first sum goes over all the nuclei of 
the problem, and the second contribution is an integral over the electronic 
cloud. This formula is exact. Let us assume however that we want to provide an empirical (and simple) expression 
for the dipole moment of the configuration considered. 
The electronic cloud around an ion in vacuum is spherical, but not in water since the solvation of the ions 
by the water molecules distorts the electronic cloud.  
As stated previously the electronic cloud can not be distributed exactly 
among the atoms of the system. However, a scheme like Atoms in Molecules\cite{bader_book} (AIM)
 provides a reasonable partitioning of the 
space. One may expect that integrating the electronic cloud around the ion (in the region assigned to the ion by
a procedure such as AIM) and adding the charge of the nucleus of the ion, one would obtain a contribution not too far away 
from +1 for the $Na^{+}$  and -1 for $Cl^{-}$. What about the water contribution?  The water molecules in contact with the ions will 
have a distorted electronic cloud, but if the solution is highly diluted most of the molecules of water 
will not be in contact with the ions, and one approximation for the contribution of the water 
molecules to the dipole moment of the entire system is to use the same charge distribution 
that provided a good dipole moment surface for pure water. 
Therefore an approximate empirical approximation for the dipole moment of a certain configuration in a diluted 
solution of NaCl in water would be:
\begin{equation}
\label{momento_dipolar_solution}
  {\bf M^{0}}
 \simeq {\bf M^{0}_{H2O}} + {\bf M^{0}_{NaCl}} 
 \simeq {\bf M^{0}_{H2O} } 
+ e ( \sum_{Na^{+}}  {\bf R_{Na^{+}} }  
  -   \sum_{Cl^{-}}  {\bf R_{Cl^{-}} } )  
\end{equation}

 In other words, we obtain  the polarization as the sum of two contributions. One due to water and the other one due to 
the ions.  This is, of course, an approximation. The dipole moment surface should be obtained from 
the electron density obtained after solving the Schr\"odinger equation. 
However the approximation described above can be regarded (for diluted solutions) as reasonable.


\textcolor{black}{ We can now focus on the PES of the salt solution. Let us assume that water-water interactions are described with 
a certain water potential model.  What to use for the ion-ion and ion-water interactions? 
 In the solid phase it has been shown\cite{JPCS_1964_25_00031,JPCS_1964_25_00045}
 that the interactions between the ions are well described by a short range repulsion plus a Coulombic interaction between 
 the ions using the charges +1 (for $Na^{+}$) and -1 ($Cl^{-}$). In fact lattice energies (and densities) are well described 
 with this approach. Let us assume that we use the same approach for the ion-ion interactions in solution. 
 The only remaining question is now : what should we use for the ion-water interactions?}
 
\textcolor{black}{ To accurately describe the property of a mixture of two components (1 and 2) 
one needs to describe correctly not only the 1-1 and 2-2 
interactions but also the 1-2 interactions. To obtain the 1-2 interactions one should use a quantum treatment.
 However quite often the 1-2 interactions are estimated by simply applying certain empirical prescriptions denoted 
 as combining rules that allows one  to estimate the 1-2 interactions once ones knows the 1-1 and 2-2 interactions. 
 For instance Lorentz-Berthelot (LB) are often used to describe the LJ interactions between different type of 
 atoms, and when they do not provide satisfactory results, deviations from LB rules are introduced.\cite{aragones_licl} 
 Concerning the Coulombic part of the potential we are quite rigid. For instance for the ion-water interaction 
 we will simply apply Coulomb law between the charges of the ions and the charges of the water model.  
 If the reader agrees with the statement  that 1-2 interactions 
 can not be obtained exactly from 1-1 and 2-2 interactions, then this idea should extend to all types of 
 contributions to the 1-2 energy (i.e short range repulsion, long range dispersion and Coulombic interactions ). 
 The hydration energy of an ion at infinite 
 dilution is mostly due to the interaction between the ion and the first hydration layers.  
 One could obtain the hydration free energy from a quantum calculation. 
 However when water is described by an empirical model, it may be the case that to 
 reproduce the hydration free energy of the ion solvated by water, the choice of +1 or -1 for the ions 
 may not be the best to reproduce {\bf simultaneously}  the hydration free energy  and the density of the solution. 
 It could be the case that reducing the charge of the ions improves the description of the hydration energies.
 Moreover, reducing the charge of the ions may improve the description of the ion-ion correlations 
 at large distances in  diluted solutions if the dielectric constant of the water model is lower 
 than the experimental value and one,  as usual, use the charges of the PES of water to describe the Coulomb 
 interactions with the ion (instead of the probably better option of using the charges of the DMS 
 of water since, in the particular case of 
 very large distances the effect of the ion on water is essentially identical to that of an electric field). 
 Reducing the charge of the ion may certainly deteriorate the description of the ion-ion interaction (so that 
 you should not use this model to describe solids or highly concentrated solutions) but it may significantly 
 improve the description of the salt solution properties at low to moderate concentrations.}

 The idea of using partial charges different from +1 and -1 ( or +ze in general) for ions is not new 
 in the literature. It has been suggested by other authors.\cite{rusos1,skinner_2014_sales}
 The idea has not been very popular, probably 
 because of the resistance to use different charges in the DMS and in the PES (i.e the "dogma"). Once you leave the "dogma", the flexibility increases. 
 Instead of  attacking this approach from the very beginning, we believe that this
 question should be decided in the battle field. The battle field in modelling is the description of 
 the properties of real systems.  Does one describe better  the experimental properties
 when using different charges to describe the PES and DMS surfaces? Do we describe better the properties
 of solutions by using charges different from +1 or -1 for the ions?  
 It should be mentioned that properties of the solution as 
 density, diffusion coefficients, vapor pressure, osmotic coefficients, chemical potentials, activity coefficients 
 depend only on the PES and not on the DMS. 

\textcolor{black}{  Certainly further work is needed to analyze this issue in detail. 
 In particular there are two problems that would be particularly useful to obtain certain conclusions. 
 The first  is the determination by computer simulation of the chemical potential and activity coefficients 
 of salts in explicit water.  Few studies have been presented so far dealing with this problem \cite{sanz07,joung_2009,joung_2013,moucka_13,panagiotopoulos_2014} and further work is certainly needed. Reproducing the Debye-Huckel limit (which is 
valid for concentrations below 0.01m) is nice but certainly not enough. For instance in the case of NaCl we must analyze the 
behavior of the activity coefficient for concentrations up to 6.14m (the solubility limit). 
 A second problem is that of determining the solubility limit of a salt in water by computer simulations. 
 Besides the technical difficulties (quite a few) it is a very hard test  for force fields as one  
 needs to simultaneously describe the salt in the solid phase
 (many salt models do 
 not even get right the melting point\cite{aragones12b}), a good description of the solvent, and a good
 description of the water-solvent interaction. No force-field so far reproduced the experimental 
 value of the solubility of 
 NaCl in water\cite{JCP_2002_117_04947,sanz07,paluch,aragones12,nezbeda_sales_1,nezbeda_sales_2} 
  (the best prediction of the solubility deviates from the experimental value by a factor of two). 
 It is clear that we have a problem. } 
 
   In agreement with the previous reasoning Kann and Skinner\cite{skinner_2014_sales}, have shown recently 
 that using partial charges smaller than +1 and -1 for the ions in  
 salt solutions it is possible to describe the variation of the diffusion coefficient 
 of water with salt concentration (increasing with concentration in the case of 
 structure breakers or chaotropes, and decreasing with concentration in the case of  structure makers or kosmotropes). 
  The key step was to leave the dogma.  Leontyev and Stuchebrukhovaa.\cite{rusos1,rusos2,rusos3} suggested 
 that to describe the PES of salt solutions, the charge of the ions should be scaled by $1/\sqrt{\epsilon_{\infty}}$ (i.e 
$\frac{1}{\sqrt{1.8}}=0.75$) . This is an interesting suggestion. In any case the charge of the ions to be used in the PES can be 
considered as an empirical parameter to be fitted to reproduce as many properties of the solution as possible. 

\textcolor{black}{ That further work is needed to analyze this  is even more obvious when one takes 
 into account that, for NaCl, no model using charges of +1 and -1 for the ions has been proposed 
 so far that describes simultaneously, the density, chemical potential 
 and melting point of the NaCl solid, the experimental values of the chemical potential of NaCl in solution up to 
 high concentrations (i.e the standard chemical potential and activity coefficients), and the solubility limit.}

\section{ Discussion  }

  We shall now discuss several issues that arise once one leaves the "dogma". 

\subsection{ The generalized hyper-surface }
 
We shall denote as surfaces  those magnitudes that depend on the positions of the nuclei only. 
The function  $E( {\bf R^{3N}},E_{el} )$ depends 
on both the position of the nuclei and of the magnitude of the external field and is a 
hyper-surface. As  was stated previously the energy of a certain configuration 
in the presence of an electric field in the z-direction can be approximated (using quantum perturbation theory) as:

\begin{equation}
\label{energia_sistema_con_campo}
 E = E^{0} - M^{0}_{z} \;\; E_{el} - \frac{1}{2} \alpha^{0}_{zz} (E_{el})^2 + ...  
\end{equation}
It is now clear that the hyper-surface E (when truncated in second order), depends on three surfaces, the PES (i.e $E^{0}$), the DMS (i.e $M^{0}_{z}$), 
and the PS (i.e  $\alpha^{0}_{zz}$ ). 
The polarization of the system in the presence of the external field is given by:

\begin{equation}
\label{momento_dipolar_sistema_con_campo}
   M_{z} =  M^{0}_{z}   + \left(   \frac{dM_{z}} {dE_{el}} \right)_{\tiny E_{el}=0} E_{el} + .. =  M^{0}_{z} +  \alpha^{0}_{zz}  E_{el} + ..
\end{equation}

 From the discussion of this paper it follows that one could use a different empirical expression to describe $E^{0}$, 
$M^{0}_{z}$ and $\alpha^{0}_{zz}$. They are three different surfaces after all. 
Let us illustrate this idea with a simple example where we use the TIP4P/2005 for the PES, 
the $\lambda$ scaling for the DMS and the Clausius-Mossoti approximation for the PS. 

\begin{equation}
\label{energia_sistema_con_campo_2}
 E =  E^{TIP4P/2005}  - \lambda \;\; M^{TIP4P/2005}_{z}  E_{el} 
    -      \frac{1}{2} \left( \sum_{j=1}^{N} \alpha_{j,zz} \right) E_{el}^2 + ...  
\end{equation}

where $\alpha_{j,zz}$ is the component $zz$ of the polarizability of molecule of water j. If one assumes that 
$\alpha$ is isotropic (a reasonable approximation for water\cite{water_polarizability}), 
and one takes the value from the gas (i.e $\alpha_{H_2O}$ ) one obtains an ever simpler expression:

\begin{equation}
\label{energia_sistema_con_campo_2}
 E =  E^{TIP4P/2005}  - \lambda \;\; M^{TIP4P/2005}_{z}  E_{el} 
    -      \frac{N}{2}   \alpha_{H_{2}O}  E_{el}^2 + ...  
\end{equation}

 This expression combines a good PES (i.e TIP4P/2005) with a much more reasonable description of the variation 
of the energy of the system with the external field (both in the linear and quadratic terms on the field).
Notice that each contribution has units of energy ( for instance in the SI, $M_{z}$ has units of C.m, $\alpha$ of $C.m^{2}/Volt$ and $E_{el}$ of 
$Volt/m$ ).  The expression of the hyper-surface when one follows the "dogma" is simply that of the previous expression 
with $\lambda=1$ and $\alpha_{H_{2}O}=0$.  It is clear that when compared to experiments the "hyper-surface" generated
when following the "dogma" is much worse than the expression we have just written, the most obvious 
consequence being an improvement in the description of the dielectric constant.

 An interesting practical remark is that if the quadratic term on the field is 
neglected  then the first order term can be written (when using the $\lambda$ scaling) 
either as $[ ( \lambda M^{TIP4P/2005}_{z} ) E_{el} ]$  
or $[$ $M^{TIP4P/2005}_{z}$ ($\lambda$ $E_{el}$ ) $]$.
That  means that if one 
uses a standard MC or MD program, where the DMS is obtained from the charges of the PES, then the results obtained
when applying an electric field $E_{el}'$ in simulations (obeying the "dogma") corresponds to those obtained when applying an 
electric field $E_{el} = E_{el}'/\lambda$ in simulations not obeying the "dogma" and using the $\lambda$ scaling. 
In general leaving the "dogma" requires rewriting the simulation program to implement two different subroutines, one 
providing the PES and another one providing the DMS. However, in case the $\lambda$ scaling approximation is used for 
the DMS, then there is no need to write the new program. Results obtained with the standard program with $E_{el}'$ correspond
to those obtained with $E_{el}=E_{el}'/\lambda$ when using the $\lambda$ scaling.

\subsection{ Electric fields and phase transitions  }

Many computer simulation programs permit the incorporation of  a static electric field (or even a dynamical one having a certain  frequency). 
No doubt many research groups will start to apply electric fields to a number
of problems and there will be dozens  of papers dealing with that.
That means that now, we should not only care about $E^{0}$, but we should seriously consider how well we represent
the changes in the energy of the system with the external field (i.e the hyper-surface). The dielectric constant is related to the
magnitude of the change in energy with the field, and for this reason it matters.
Leaving the "dogma" will provide a better description of the hyper-surface so that predictions will be more reliable.

Another  interesting issue to consider in the future is the effect of electric fields on phase transitions. 
This is even more complex and challenging.  Now what matters is the difference in the value of the dielectric 
constant between the two phases. \textcolor{black}{ In the particular case of the fluid-solid transition, the most common scenario in 
the case of molecular polar systems is  that of a solid with a low dielectric 
constant (the  constraints imposed by the lattice will not allow large fluctuations 
of the total polarization of the solid so that the dielectric constant will be small) and a liquid with a moderate dielectric 
constant. If the prediction of the model for the dielectric constant of the liquid phase is  good, then the simulations will predict 
(correctly) a decrease in the melting point due to the presence of the electric field. However there is an important 
exception to this common scenario : the ice Ih-water transition. } 
Experimentally the dielectric constant of ice Ih at the melting point is slightly larger than that of water (the existence 
of proton disorder in the solid\cite{pauling35} allows ice Ih  to response efficiently to an electric field).  
According 
to this when applying an electric field the melting 
point will increase  as the polarization of ice Ih is larger 
than that of water and becomes further stabilized by the electric field. 
What will happen in simulations?  Although the interest in the dielectric constant of ice Ih has been rather small 
recently it has become clear that within the formalism of the "dogma" the dielectric constant of ice Ih 
is lower than that of liquid water (by about 15 per cent  for TIP4P models, and by about 50 per cent  for models  
such as TIP3P, SPC/E or TIP5P).  Therefore these models, within the formalism of the "dogma", 
will predict (incorrectly) that the melting point of ice Ih decreases by a small amount (TIP4P like models) and significantly 
(  TIP3P, SPC/E and TIP5P ).  For huge electric fields (of about 1V/nm) 
a ferroelectric  Ih (or Ic) phase\cite{ferroelectric_slater,ferroelectric_loerting} will be stabilized and 
one should expect a huge increase in the melting point with the field, as has been observed recently by 
Yan, Overduin and Patey.\cite{patey_14} This is interesting and probably relevant for water under confinement, but  not
so important for bulk water because the intensity of the electric field required  to stabilize the ferroelectric field is 
huge and is beyond the electric breakdown point of water\cite{electric_breakdown} (which is of about 0.01 V/nm).
Therefore for experimental studies of bulk water  the key variable to understand the impact of an electric field on 
the ice Ih - water phase transition  is the difference between the dielectric constant of these two phases.\cite{aragones11b}
If the $\lambda$ scaling is used for TIP4P/2005 (with the same value of $\lambda$ for ice Ih and for the liquid phase) 
the description of the dielectric constant of both phases improves but  
the dielectric constant of ice Ih is still slightly lower than that of water. 
 In order to make predictions that can be compared to experiments
it is necessary to use different values of $\lambda$ for ice Ih and for the liquid phase as we did in previous work\cite{aragones11b}.
  A similar problem was faced  by Skinner and co-workers 
to describe the dielectric constants of ice Ih and water. These authors use the E3B model as the PES for both 
the liquid and the solid phase. However, to describe the DMS they use a polarizable model with different 
parameters for the solid and liquid phases; therefore   the experimental values of $\epsilon_{r}$ were reproduced for both 
phases. 
If one does that,  the predictions for the effect of the electric field on the phase transition would make sense and 
could be compared to experimental results. 
Now that  interest in  the effect of electric fields in phase transitions is growing,
the issue of the dielectric constant of the two phases involved matters, and the idea of using different charges ( or 
even empirical expressions ) to describe the PES and the DMS may be useful. 

 The idea of using different charges and/or methodologies in different phases to obtain the DMS  
is fine for determining the properties of each phase, or the effect of an electric field on  a phase
transitions.  However, this approach can not deal with problems like 
interfacial properties or nucleation phenomena since it is not clear how to incorporate interfacial molecules 
(which are not neither fully liquid nor fully solid ) into the treatment. 
Polarizable models (and/or ab initio calculations) in principle do not have this problem as these methods provide
a DMS that can be used for both phases. Whether these treatments are able to describe quantitatively the 
dielectric constant of both phases need to be analyzed in more detail although recent results suggest 
that this may indeed be the case.\cite{epsilon_dft,wang13}
In any  case the possibility of using a non-polarizable model for the PES and a   polarizable model for the 
DMS is also open. 

 This paper does not pretend to be a heroic defense of non-polarizable models. These models have limitations, as it 
is clear from the water test. Rather this paper  advocates that the discussion about the quality of water 
models to describe the dielectric constant was probably  wrong, because it was based on the assumption that
the same charges should be used to describe the PES and the DMS. This is not necessary.
Probably it is not in the prediction of the 
dielectric constant where polarizable models defeat clearly non-polarizable models.  
It is rather for properties like the vapor pressure, cluster properties, critical pressure, second virial
coefficient, and vaporization enthalpy where polarizable models show their superiority over non-polarizable 
ones.\cite{wang13,baranyai13,kolafa_polarizable} 
Certainly, everything suggests that  models with parameters depending on the local environment (i.e polarizable) provide
a better PES (especially when the model is used to describe properties of the gas and of condensed phases simultaneously). 

\subsection{  Classical electrostatics is not quantum mechanics.  }

\textcolor{black} { The dipole moment of a configuration can be easily obtained, once the positions of the nuclei are
provided and the electron density is known. 
The formula used in quantum mechanics to obtain the dipole moment of a certain configuration is identical
to the formula used in classical electrostatics to obtain the dipole moment of a certain distribution of 
point charges and a continuous charge distribution. 
Thus, concerning the DMS, classical electrostatics and quantum mechanics get along very well. 
What about the energy, i.e the PES? 
 It is instructive to write the expression of the energy as 
obtained from DFT\cite{review_aimd}  (in the absence of an electric field ): }

\begin{equation}
E^{0} ({\bf R^{3N}})= 
E[ \rho^{0}( {\bf r};{\bf R^{3N}})] =    \frac{1}{2} \int \int \frac{ \rho^{0}( {\bf r_{1}} ) \rho^{0}( {\bf r_{2}} ) }
{ | {\bf r_{1}}  - {\bf r_{2}} | }    d{\bf r_{1}}  d{\bf r_{2}} - \sum_{\gamma=1}^{3N} \int  \frac{Z_{\gamma}} { | {\bf  r} - {\bf R_{\gamma}}  | } \rho^{0}({\bf r})  d{\bf r}  + \sum_{\gamma} \sum_{\eta>\gamma}  \frac{ Z_{\gamma} Z_{\eta} } { R_{\gamma\eta} }
\nonumber 
\end{equation}

\begin{equation}
 \frac{-1}{2} \sum_{i=1}^{n_{e}} \int \Psi_{i}({\bf r})
  \nabla^{2}\Psi_{i}({\bf r}) d{\bf r} + E_{XC}[ \rho^{0}( {\bf r}) ]
\end{equation}


 

%

\textcolor{black}{ where the electronic density at point ${\bf r}$ has been approximated 
by the sum of the contributions of different orbitals $\Psi_{i}$,
i.e, $  \rho^{0} ( {\bf r} ) = \sum_{i=1}^{n_{e}} | \Psi_{i}({\bf r}) |^{2} $ and we used atomic units. 
For each configuration of the nuclei, the electron density will be obtained by minimizing the energy of the 
system with respect to the electron density. 
In the functional the first three terms have a simple electrostatic origin, namely
the repulsion energy between the electronic clouds, the attractive energy between the nuclei and the electronic
cloud,  and the repulsion energy between the nuclei. These terms can be easily understood from pure electrostatics.  Let us now analyze the last two terms. 
One is the kinetic energy of the electrons, and the last one,  $ E_{XC}$, represents the 
exchange correlation functional. These two terms can  not be derived from classical electrostatics. 
Empirical potentials recognize that and this is the reason why LJ centers are often included to incorporate 
long range dispersive  forces and short range repulsive forces as an implicit way of including part of the contribution of 
the $ E_{XC}$ and kinetic energy terms.
One should not forget 
that  the exact  energy of a configuration can not be obtained from simple
formulas from electrostatics and/or from any treatment based on an analogous electrostatic problem. The presence 
of the exchange correlation and kinetic energy terms is the reason why the quantum world 
can not be mapped into a problem of classical 
electrostatics. Thus concerning the energy (and the electron density, which will be obtained from minimization 
of the functional) the classical electrostatics and the quantum chemistry are divorced. They simply predict different 
things, because they are using different functionals. The laws of quantum chemistry can not be mapped exactly into an 
analogous electrostatic problem. 
One may think that using the same electrostatics entitities (partial charges, diffusive charges, fixed dipoles, 
induced dipoles, quadrupoles ..) to describe the PES and the DMS is a sign of consistency.
Using the same charges for the PES and the 
DMS is consistent in an imaginary world where the interaction between molecules is given by  LJ centers
and charges and/or multipoles that obey a certain simple model derived from classical electrostatics.
However, nature follows the laws of quantum chemistry.
Once one recognizes that classical electrostatics can not describe the PES and the DMS simultaneously, the
step to use different models to describe the PES and the DMS follows naturally. 
In fact, we have already mentioned that within team D (i.e analytical potentials) there are two groups:
analytical ab-initio potentials and empirical potentials. Interestingly, the community developing ab-initio
analytical potentials  is open to the use of different  fits for the PES and the 
DMS\cite{neural_network_1,neural_network_2,neural_network_3}, and they regard
charges, or partial charges, as merely fitting parameters to surfaces that were obtained from high level ab initio
calculations. They simply want to reproduce the high level results for the two surfaces 
with high accuracy and they do not attach  so much physical significance  to the fitting parameters. However in 
the community developing empirical potentials, we replaced the quantum problem by a simple electrostatic problem, and 
then implicitly assumed that "for consistency" the same charges should be used for the PES and the DMS. 
Using simple classical electrostatic models for the energy is fully inconsistent 
with the laws of the microscopic world. For this reason 
we do not see any reason  when developing empirical potentials (with parameters obtained to reproduce 
experimental properties) why we could not use different models/treatments for the PES and the DMS. }

  \subsection{ Coarse grained models of water and the dielectric constant }

 In 1990 Tomas Boublik, on a sabbatical leave in Madrid, taught me (among many other things) 
the perturbation theory proposed by  Wertheim\cite{wertheim} for 
associating fluids, that was further extended by Chapman, Jackson and Gubbins\cite{chapman_jackson_gubbins}, and which is nowadays known as SAFT.\cite{primer_paper_con_nombre_saft}
He  figured out that the theory could be very useful to implement an equation of state for fused hard sphere
chains\cite{boublik90}, extending his classical work on the equation of state of hard-convex bodies\cite{boublik_nezbeda_review} and hard spheres mixtures.\cite{cs_boublik} 
In Wertheim's/SAFT  theory the molecules are described by strong short range associating sites (emulating the hydrogen bond) 
and the properties can be computed by using well defined approximations. The theory is becoming quite successful
for practical applications. Quite often no dipoles or partial charges are used to define the interactions between 
molecules. Probably it is fair to say that it is one of the most popular  perturbation theories of liquids after van der 
Waals.\cite{GrayGubbins_volume_2}  
In this theory  water is described as a spherical molecule, with four short range association 
sites, two hydrogens and two "lone pair electron" sites located in a tetrahedral arrangement. A very successful
model within this framework is the Kolafa-Nezbeda model of water\cite{kolafa_nezbeda_model_1,kolafa_nezbeda_model_2,vega_monson_98}
composed by just a hard sphere and four association sites. 
The contribution of dispersive forces to the properties can be obtained either using a mean field approach, or 
eventually modifying the Kolafa Nezbeda model so that one has a LJ center plus four associating sites.\cite{lourdes_water,MP_2006_104_3561} This
is a reasonable model of water, and  has been shown that when used in combination with SAFT it can describe
many properties of water. Thus SAFT provides a good description of water because the potential used to describe 
water (i.e the PES) is simple but still reasonable.  Probably, in a water test like that one presented in Table \ref{score}
these models will obtain a score lower than TIP4P/2005 but probably not worse than TIP3P (i.e $2.7$). 
The same is true for the mW model of water of Molinero and co-workers. In this case, the tetrahedral coordination of water is induced, not by using
associating sites as in SAFT's approach, but by introducing three body forces\cite{weber_stillinger}. 
This model has no charges. Molinero and co-workers have implemented the water test (not for all the properties considered
in our initial test but for some of them) and showed that mW describes reasonably well water.\cite{mw_water_test}. The score was lower
($6.1$) than TIP4P/2005 ($7.8$ for the properties selected by Molinero and co-workers) but still reasonable. Thus mW is a reasonable PES of water. 
Let us emphasize again that both the SAFT and the mW PES do not use partial charges
and still provide a reasonable description of water. Partial charges are certainly a possibility to induce tetrahedral
order in water, but it is clear that it is not the only one. It is clear that 
the PES of water should indeed favor tetrahedral coordination of the molecules. 
 
 Now let us state a common criticism received by these models: " they are not real models of water  
since they have no partial charges and therefore their dielectric constant is 1 ". At this point I hope to have succeeded 
in convincing the reader that this statement is absolutely wrong. It is based on the "dogma", i.e on the 
implicit assumption that the charges used to describe the PES  should also be used in the description of the DMS.
In this case, there are no charges in the PES, but you could certainly use charges to describe the DMS. 
I do not see any reason, why these type of "coarse grained" models could not be used for modelling salt solutions. In fact some 
attempts to do that have been undertaken in the past\cite{JPCB_1999_103_10272,saft_sales,mw_nacl} by introducing 
a short range attraction to describe the interaction between the ions and water, and by using a Yukawa like potential 
to describe the ion-ion interactions.

\section { Conclusions}
  The main conclusion of this work is simple.
 For water there are two surfaces, the PES and the DMS (strictly speaking three if one includes in the 
treatment the PS surface). Empirical potentials are aimed at describing  the 
 PES (i.e the energy of the system in the absence of the field).  It is also possible to use 
 empirical expressions to describe the DMS. 
 In the case  you use partial charges/multipoles to describe
 the PES  this is fine but there is no reason to use the same partial charges/multipoles to describe the DMS. 
 If you do not use partial charges/multipoles in the description of the PES, as in coarse grained models,  
 there is no reason why you could not use partial charges/multipoles in the description of the DMS. 
 The implicit assumption that the same charges should be used in the description of the PES and of the DMS is a "dogma".
 This "dogma" has contaminated all our analysis about the ability of water models to describe the dielectric constant. 
 We need to revise our thinking about this property.
 There is nothing wrong (neither physically nor from a practical point of view) in 
 using different charges for the PES and for the DMS. Therefore the charges used for the PES are not necessarily the 
 best to describe the DMS. The error also goes the other way around. In cases where 
 the charges to be used in the description of the DMS seems more or less obvious (as when you have ions)   
 these charges may not necessarily be the best to describe the PES. The idea also extends to the  
 polarization surface PS. The charges used to describe the PES and/or the DMS do not provide any information 
 about how the polarization of the system changes with an electric field. For this reason  it is also possible
 to include an approximate empirical expression to describe the PS. 

   Since we are not solving the Schr\"odinger equation let us be practical when describing the PES and the DMS. 
 Empirical potentials should provide a good PES, thus describing all properties of water in the absence of 
 an electric field. Once you have 
 a good PES, then you need a good DMS to describe the dielectric constant of water. 
 If the descriptions of the PES and DMS are correct 
 then you will correctly describe all the experimental properties of water, including the dielectric constant. 
   Thus, the conclusion, is that water is one molecule, with two surfaces (three when the PS is included), 
and that we have been doing during years of water simulations, one mistake. 

    Of course, although we  used water for the discussion, since this is the molecule we  have studied in more detail
 during  these years (and it is probably the molecule that has been studied by more people) the central idea of 
 this paper can also be extended to other molecules. The PES, the DMS and the PS are three  surfaces that 
 should be fitted using different parameters. Now that work aimed to study the effect of electric fields on 
 matter are appearing in the literature, a good PES, a good DMS and (to lesser extent) a good 
 PS are needed. 
 If  we continue using the "dogma" to describe the PES, DMS and PS surfaces, then the predictions from computer simulations on the 
 impact of electric fields on the properties and on the phase transitions of water (particularly on the ice Ih-water transition) may be
 incorrect. 

   It is time to depart from the path initiated by Bernal and Fowler.\cite{bernal33} 

\section{acknowledgments}
This work was funded by grant FIS2013/43209-P of the Spanish Ministery of Education.
The initial idea for this manuscripts came after an interesting
discussion with Prof. Dellago and Tobias Morawietz in Wien,
October 2013, on the use of neural networks on quantum chemistry
after the interesting thesis of Dr. Geiger devoted to the use of
neural networks as order parameters.We thank Prof. Skinner, Prof.
Molinero, Prof. Hamm, Prof. Galli and Prof. Panagiotopoulos for
sending me preprints of their work prior to publication. Helpful
discussions with Prof. Abascal, Prof. Skinner and Prof. Paesani
are gratefully acknowledged. We thank Dr. McBride, Dr. Noya,
Prof. Skinner, Prof. Benavides , Prof. Abascal and J. Rene for
a critical reading of the manuscript. We would like to dedicate
this paper to the memory of Tomas Boublik who passed away in
January 2013.We would also like to congratulate Ivo Nezbeda on
the occasion of his 70th anniversary. Tomas and Ivo, along many
others developed the Czech School on Statistical Mechanics, that
generated a lot of good science even in difficult times.



\end{document}